\documentclass[preprint,12pt]{elsarticle}
\usepackage{amssymb}
\usepackage{amsmath}
\newtheorem{theorem}{Theorem}
\newtheorem{corol}{Corollary}
\newtheorem{lemma}{Lemma}
\newtheorem{prob}{Problem}
\newtheorem{defi}{Definition}
\newtheorem{example}{Example}

\journal{Journal Name}

\begin{document}
	
	\begin{frontmatter}
		
		
		
		\title{Exact Volume of Zonotopes Generated by a Matrix Pair
		\thanks{Work supported by the National Natural Science Foundation of China (Grant No. 61273005)}}
		
		
		\author{Mingwang Zhao}
		
		\address{Information Science and Engineering School, Wuhan University of Science and Technology, Wuhan, Hubei, 430081, China \\
			Tel.: +86-27-68863897\\
		Work supported by the National Natural Science Foundation of China (Grant No. 61273005)}
		%	\email{zhaomingwang@wust.edu.cn} 
		
		\begin{abstract}
			In this article, we define a class of special zonotopes
			generated by a matrix pair with finite-interval
			parameters. We discuss the relationship between the volume of these
			zonotopes and the controllability of one aspect (the volume of the
			controllable region) of the dynamic systems. We present a corollary
			and develop an effective recursive method to compute the volume of the special
			zonotopes.
			Furthermore, we develop two recursive and analytical volume-computation
			methods for the finite- and infinite-time controllable regions with real eigenvalues. We conduct numerical experiments to demonstrate the
			effectiveness of these new volume-computation methods for
			zonotopes and regions.
		\end{abstract}
		
		
	\end{frontmatter}
	
%	\linenumbers
	
	\section{Introduction}
	\label{S:1}
	
	In control theory and engineering, linear dynamic systems in the
	discrete-time case can be formulated as follows:
	
	\begin{equation}
	x_{k+1}=Ax_{k}+Bu_{k}, \quad x_{k} \in R^{n},u_{k} \in R^{r}, \label{eq:a01}
	\end{equation}
	\noindent where $x_{k}$ and $u_{k}$ are the state variable and input
	variable, respectively, and matrices $A \in R^{n \times n}$ and $B \in
	R^{n \times r}$ are the state matrix and input matrix, respectively,
	in the system models \cite{Kailath1980},\cite{Chen1998}. 
	To investigate the controllability of the
	linear dynamic systems \eqref{eq:a01}, the input variables $ u_{k} $ are needed to be bounded and normalized  for the following reasons. 
	
	1) The many practical controlled plants are with the bonunded input variables or the input saturation elements, that is, the input variables $ u_{k} $ are bounded;
	
	2) To compare properly the state control ability of the input variables between the different systems or in the one system with the different parameters in system models $\{A,B\} $, the state variables  and  the input variables  of these systems are with the matching scale and normalization, respectively.
	
	Therefor, in this paper,  the state variables between the different systems are with the matching scale, and the input variables $ u_{k} $ are bounded and normailized as 	$ \Vert u_{k} \Vert_{ \infty} \leq1$. Then, the $N$-steps controllable
	region $R_{c,N}$ and reachable region $R_{r,N}$ of the  systems \eqref{eq:a01}  can be defined as	
	
	\begin{align}
		R_{c,N}
		& = \left \{x_{0} | x_{0}=- \left(A^{-N}P_{N} \right)U_{N}, \left \Vert U_{N} \right \Vert _{ \infty} \leq1 \right \} \notag \\
		& = \left \{ x_{0} | x_{0}= \left(A^{-N}P_{N} \right)U_{N}, \left \Vert U_{N} \right \Vert _{ \infty} \leq1 \right \} \label{eq:a02} \\
		R_{r,N}
		& = \left \{ x_{N} | x_{N}=P_{N}U_{N}, \left \Vert U_{N} \right \Vert _{ \infty} \leq1 \right \} \label{eq:a03}
	\end{align}
	\noindent where $x_{0}$ and $x_{N}$ are respectively the initial and terminal
	states of the dynamic systems in the control process, 
	$U^T_{N}= \left[
	u_{N-1}^{T}, u_{N-2}^{T}, \cdots, u_{0}^{T} \right]$ is the control input sequence, and
	$P_{N}= \left[B,AB, \cdots,A^{N-1}B \right]$ is the controllable
	matrix \cite{Lin1970} \cite{Fisher1987} \cite{Lindner1989} \cite{Lasserre1991} \cite{Lasserre1993} \cite{Hulin2001} \cite{HuMiQiu2002} \cite{KURVARA:07} \cite{KOSTOUSOVAE2009}. Because the controllable
	region and reachable region defined as above can be transformed each other, without loss of generality, only the reachable region $	R_{r,N}$ and the reach ability are discussed later and the obtained conclusions can be generlized conveniently to the controlable region $	R_{c,N}$ and the control ability.
	
%	出于以下原因，可以认为;上述定义的状态能控/能达域完整地反映了线性离散定常系统\eqref{eq:a01} 的输入变量对状态变量的控制/到达能力。

%1) $N$步状态能控域$R_{c,N}$ /能达域$	R_{r,N}$ 越大,则系统在$N$步之内的能控/能达的状态就越多;

%2) 对系统状态空间的同一个状态$x_1$,$N$步状态能控域$R_{c,N}$ /能达域$	R_{r,N}$ 越大,则将状态$x_1$控制到空间原点(或从空间原点控制到状态$x_1$),存在着控制时间更少、燃料总消耗($\Sigma_k \; \Sigma_i \; \vert u_{k,i} \vert )$ 更少的控制策略；

%3) 对系统状态空间的同一个状态$x_1$,$N$步状态能控域$R_{c,N}$ /能达域$	R_{r,N}$ 越大,则将状态$x_1$控制到空间原点(或从空间原点控制到状态$x_1$),存在着更多的控制策略。

%对式\eqref{eq:a03}定义的能达域,我们可知：
	
	Based on the definition of the reachable region  by Eq. \eqref{eq:a03}, we know,	
	
%	1) 能达域$R_{r,N}$ 越大(例如,$R^{(1)}_{r,N} \subset R^{(2)}_{r,N} $),则系统在$N$步之内能达的状态就越多,能达的范围就越大;
	
	1) The larger the size of the reachable region $R_{r,N}$ is (e.g., $R^{(1)}_{r,N} \subset R^{(2)}_{r,N} $), the more the reachable states of the systems in $N$-steps are, and the larger the reachable range in the state space is;
	
%	2) 对将系统状态空间原点的状态控制到给定的同一个状态$x_1$的控制问题,能达域$	R_{r,N}$ 越大,
	
	2) For the reaching control problem that the state is controlled from the orign in the state space to the given same state $x_1$, if the size of the reachable region $R_{r,N}$ is larger, 
	
	2.1) there exists a control strategy with the less control time and the faster response speed;
	 
%	2.1) 存在着控制时间更短的控制策略,即存在更快到达控制目标、响应更快的控制策略；
	
	2.2) there exist more control strategies, that is,  the larger the size of the solution space of the input squence  for the reaching control proble and then the the easier to design and implement for the reaching control systems.
		
%	2.2) 存在着更多的控制策略,即输入变量的控制序列的解空间更大,控制更容易实现。
	
%因此,可以认为:上述定义的能达域的大小很好地反映了线性离散定常系统\eqref{eq:a01} 的输入变量对状态变量的到达控制能力。从几何分析的角度,控制领域的能达域实际上可视为$n$维状态空间中的一个几何体,其大小可由该几何体的几何形状和体积来表征。在形状相似或相近情况小,体积越大则反映出该系统的输入变量对状态变量的到达控制能力就越强。文[]讨论了能达域的顶点、边界的确定或近似(逼近)计算。本文主要讨论能达域的体积的精确计算。

Therefore, it follows that the size of the reachable region can  reflects well the state-reaching control ability of the input variables of the linear time-invariant discrete-time systems \eqref{eq:a01}. From the perspective of geometric analysis, in fact, the reachable region in control theorey and engineering field can be regarded as a geometry in $n$-dimension state space and can be characterized by its surface, shape and volume. When the geometry shapes are same or approximate, the larger the geometry volume are, the larger the geometry size. 

%The vertexes in the surface of the reachable region are discussed and computed in paper [], the volume and shape are discussed and computed conveniently in this paper.

	To accurately measure the controllability of the systems,
	the volumes of the regions $R_{c,N}$ and $R_{r,N}$ must be
	computed. Based on volume computing, the controllability can be
	optimized and then the control performance of the closed-loop
	control systems for the open-loop systems \eqref{eq:a01} can be
	prompted.
	
	In fact, the regions defined in eqs. \eqref{eq:a02} and
	\eqref{eq:a03} can be considered a class of zonotopes spanned by a vector set with a parameter set in the finite interval. These zonotopes can be defined as
	follows \cite{McMu1971} \cite{GovKri2010} \cite{BecRob2015}.
	\begin{defi} \label{d10} The zonotopes spanned by the $n$-dimensional ($n$-D) vectors of matrix $Z_{m}=[z_{1},z_{2}, \dots,z_{m}] \in R^{n \times m}$ and the parameter set with a finite interval are defined as
		\begin{equation} \label{eq:a04}
		C_{q}(Z_{m}) = \left \{ \left. \sum_{i=1}^{m} c_{i}z_{i} \right| \forall c_{i} \in[0,1], i= \overline{1,m} \right \}
		\end{equation}
		\noindent where $q= \textrm{rank}(A_{m}), c_{i} \left(i=
		\overline{1,m} \right)$ are the parameters representing the
		zonotope, and vectors $z_i(i= \overline{1,m})$ are called as the generators of the zonotopes. The zonotopes are the $q$-D parallel polytopes in 
		the $n$-D space, and they are convex.
	\end{defi}
	
	Similar to the above definition of zonotope $C_{q}(Z_{m})$, as eqs. \eqref{eq:a02} and \eqref{eq:a03} describe the controllability region $R_{c,N}$ and reachability region $R_{r,N}$, we can define a new type of zonotope generated by the matrix pair $ \{A,B \}$ as follows.
	
	\begin{defi} \label{d101} The zonotopes generated by the matrix pair $ \{A,B \}$ and the parameter set with a finite interval are defined as
		\begin{equation}
		E_{q}(P_{N}) = \left \{ \left. \sum_{i=1}^{rN} c_{i}p_{i} \right| \forall c_{i} \in[0,1], i= \overline{1,rN} \right \} \label{eq:a05}
		\end{equation}
		\noindent where $A \in R^{n \times n}$ and $B \in
		R^{n \times r}$, $P_{N}= \left[B,AB, \cdots,A^{N-1}B \right]= \left[p_1,p_2, \cdots,p_{rN} \right]$, $q= \textrm{rank}(P_{N})$, $c_{i} \left(i=
		\overline{1,rN} \right)$ are the parameters representing the
		zonotope, and the matrix pair $ \{A,B \}$ is called the generator pair of the zonotopes.
	\end{defi}

	It follows from the definition of the zonotopes $E_{q}(P_{N})$ that
	the regions $R_{c,N}$ and $R_{r,N}$ can be gotten from the zonotope $E_q(P_N)$ by some linear transformations \cite{GirGue2008} \cite{AltKro2011}. Since the
	controllability of the dynamic systems is related to these regions, and the geometric volume is a key index for investigating these regions, to investigate the controllability can in some respects be carried
	out by investigating the volumes of these zonotopes.
	
	In fact, the exact volume of the zonotope $C_{q}(Z_{m})$ generated by $m$ vectors $z_i(i= \overline{1,m})$ can be computed as the sum of the determinants of any $n$ vectors from the vectors $z_i(i= \overline{1,m})$.
	These relevant results can be summarized in the
	following theorem \cite{DyerGerHuf1998} \cite{GovKri2010}.
	
	\begin{theorem} \label{t10}
		For any full row rank matrix $Z_{m} \in R^{n \times m}$, the volume of the $n$-D zonotope $C_{n}(Z_{m})$ spanned by the vectors of $Z_{m}$ can be computed as
		\begin{equation} \label{eq:a06}
		V_{n} \left(C_{n}(Z_{m}) \right)= \sum_{(i_{1},i_{2}, \dots,i_{n}) \in \Omega_{m}^{n}} \left| \det \Lambda_{i_{1}i_{2}...i_{n}} \right|
		\end{equation}
		\noindent where $ \Lambda_{i_{1}i_{2}...i_{n}}= \left[z_{i_{1}},z_{i_{2}}, \cdots,z_{i_{n}} \right]$, and the column-label $n$-tuple set $ \Omega_{m}^{n}$ consists of all possible $n$-tuples $(i_{1},i_{2}, \dots,i_{n})$ whose elements are picked from the set $ \left \{ 1,2, \cdots,m \right \}$ and are sorted by their values. The computational complexity of the volume-computation method, \textit{i.e.}, the times computing the $n \times n$ determinant values, is
		\begin{equation} \label{eq:a07}
		\frac{m!}{ \left(m-n \right)!n!}
		\end{equation}
		times, noted as the polynomial time $ \mathcal{O}(m^{n})$ on the vector number $m$.	
	\end{theorem}
	
	The volume of the zonotope $E_q(P_N)$ generated by the matrix pair $ \{A,B \}$, as computed by eq. \eqref{eq:a06}, will have complexity $ \mathcal{O}((rN)^{n})$, \textit{i.e.}, the complexity will be $ \mathcal{O}({N}^{n})$ on the time variable $N$. For many practical problems in control theory and engineering, the dimensions $n$ and $r$ in the matrix pair $ \{A,B \}$ are finite, but the sampling-step number $N$ is a time variable that will gradually increase. Considering that $N$ is gradually increasing and even will approach infinity, the focus of the computational complexity for the zonotope $E_q(P_N)$ is on the time variable $N$ but not the finite dimension variables $n$ and $r$. Therefore, we
	focus on the following two methods to compute the volume.
	
	\begin{prob} \label{p10}
		The exact volume computation of the finite-time zonotope $E_q(P_N)$ generated by the matrix pair $ \{A,B \}$ with the lower complexity on time variable $N$.
	\end{prob}
	
	\begin{prob} \label{p11}
		The analytically exact volume computation of the infinite-time zonotope $E_q(P_ \infty)$ generated by the matrix pair $ \{A,B \}$ with complexity $ \mathcal{O}(1)$.	
	\end{prob}
	
	In this paper, first, for \textbf{Problem \ref{p10}}, the recursive computation volume of the finite-time zonotope $E_q(P_N)$ with  the general matrix pair $\{A,B\}$ will be discussed in section 2, and a new computation method with complexity $ \mathcal{O}({N}^{n-1})$ is obtained. In section 3, the same problem for the matrix $A$ with $n$ real eigenvalues will be discussed, and a new computation method with complexity $ \mathcal{O}({N})$ is proposed and proven. For \textbf{Problem \ref{p11}}, the analytic computation method for infinite $N$ and $n$ real eigenvalues will be given and proven with complexity $ \mathcal{O}(1)$ in section 4. Finally, the numerical experiments for the computation methods proposed in this paper will be carried out in section 5. The effective computation methods for the zonotope $E_q(P_N)$ when the matrix $A$ has a more complex eigenvalue distribution than $n$ real eigenvalues will be investigated in future work.
	
	\section{Volume Computation of Zonotope Generated by Matrix Pair}
	
	As mentioned earlier, the regions $R_{c,N}$ and $R_{r_N}$ can be represented essentially by the zonotopes $E_{n}(A^{-N}P_{N})$ and $E_{n}(P_{N})$, respectively.
	Therefore, based on \textbf{Theorem \ref{t10}}, the volumes of these regions can be computed conveniently by computation of these zonotope volumes. So, we have
	\begin{align}
		V_{n} \left(R_{c,N} \right) & = \left| \det A \right|^{-N} V_{n} \left(R_{r,N} \right) \label{eq:a08} \\
		V_{n} \left(R_{r,N} \right) & = 2^n V_{n} (E_{n}(P_{N})) \label{eq:a09}
	\end{align}
	Hence, only the volume computation of the zonotope $E_{n}(P_{N})$ is studied in detail.
	
	\subsection{recursive computation method}
	
	From \textbf{Theorem \ref{t10}}, we have the following corollary on the volume computation for the special zonotope $E_{n}(P_{N})$ generated by the matrix pair $ \{A,B \} $.
	
	\begin{corol} \label{c12}
		For any matrices $A \in R^{n \times n}$ and $B \in R^{n \times r}$, the volume of the zonotope $E_{n}(P_{N})$ generated by pair $ \{A,B \}$ can be computed recursively by the following equation with computational complexity $ \mathcal{O}(N^{n-1})$ on time variable $N$:
		\begin{align}
			V_{n} (E_{n}(P_{N}))
			& = \left(1+ \left| \det A \right| \right)V_{n}(E_{n}(P_{N-1}))- \left| \det A \right|V_{n}(E_{n}(P_{N-2})) \notag \\
			& \quad+ \sum_{j=1}^{r} \sum_{k=1}^{r} \sum_{(i_{1},i_{2}, \cdots,i_{n}) \in
				\varTheta_{0,0}^{j} \times \varTheta_{1,N-2}^{n-j-k} \times \varTheta_{N-1,N-1}^{k} }
			\left| \det \varPsi_{i_{1}i_{2} \cdots i_{n}} \right| \label{eq:a10}
		\end{align}
		\noindent where $ \varPsi_{i_{1}i_{2} \cdots i_{n}}=[p_{i_{1}},p_{i_{2}}, \cdots,p_{i_{n}}]$, $p_{i}(i= \overline{1,rN})$ is the $i$-th vector of matrix $P_{N}$,and the $j$-tuple set $ \varTheta_{N,M}^{j}$ consists of all possible $j$-tuples
		$(i_{1},i_{2}, \dots,i_{j})$
		whose elements are picked from the set
		$ \left \{ rN+1,rN+2, \cdots,r(M+1) \right \}$
		and sorted by the values
		$$ \varTheta_{*}^{j} \times \varTheta_{*}^{m-j}= \{ \left.(i_{1},i_{2}, \cdots,i_{m}) \right| \forall (i_{1}, \cdots,i_{j}) \in \varTheta_{*}^{j},
		\forall (i_{j+1}, \cdots,i_{m}) \in \varTheta_{*}^{m-j}
		\} $$
	\end{corol}
	
	\textbf{Proof}. (1) By the volume-computation equation \eqref{eq:a06}, we have
	\begin{align*}
		V_{n} (E_{n}(P_{N})) &
		= \sum_{(i_{1},i_{2}, \cdots,i_{n}) \in \varTheta_{0,N-1}^{n}} \left| \det \varPsi_{i_{1} i_{2} \cdots i_{n}} \right| \\
		&
		= \left \{ \sum_{(i_{1}, \cdots,i_{n}) \in \varTheta_{0,N-2}^{n}}
		+ \sum_{(i_{1}, \cdots,i_{n}) \in \varTheta_{1,N-1}^{n}}
		- \sum_{(i_{1}, \cdots,i_{n}) \in \varTheta_{1,N-2}^{n}} \right. \\
		& \left.
		\quad+ \sum_{j=1}^{r} \sum_{k=1}^{r} \sum_{(i_{1}, \cdots,i_{n}) \in
			\varTheta_{0,0}^{j} \times \varTheta_{1,N-2}^{n-j-k} \times \varTheta_{N-1,N-1}^{k} }
		\right \}
		\left| \det \varPsi_{i_{1} i_{2} \cdots i_{n}} \right| \\
		& =V_{n}(E_{n}(P_{N-1}))+ \left| \det A \right| \sum_{(i_{1}, \cdots,i_{n}) \in \varTheta_{0,N-2}^{n}} \left| \det \varPsi_{i_{1} i_{2} \cdots i_{n}}
		\right| \\
		& \quad- \left| \det A \right| \sum_{(i_{1},i_{2}, \cdots,i_{n}) \in \varTheta_{0,N-3}^{n}} \left| \det \varPsi_{i_{1} i_{2} \cdots i_{n}} \right| \\
		& \quad+ \sum_{j=1}^{r} \sum_{k=1}^{r} \sum_{ (i_{1},i_{2}, \cdots,i_{n}) \in
			\varTheta_{0,0}^{j} \times \varTheta_{1,N-2}^{n-j-k} \times \varTheta_{N-1,N-1}^{k} }
		\left| \det \varPsi_{i_{1} i_{2} \cdots i_{n}} \right| \\
		& = \left(1+ \left| \det A \right| \right)V_{n}(E_{n}(P_{N-1}))- \left| \det A \right|V_{n}(E_{n}(P_{N-2})) \\
		& \quad+ \sum_{j=1}^{r} \sum_{k=1}^{r} \sum_{(i_{1},i_{2}, \cdots,i_{n}) \in
			\varTheta_{0,0}^{j} \times \varTheta_{1,N-2}^{n-j-k} \times \varTheta_{N-1,N-1}^{k} }
		\left| \det \varPsi_{i_{1} i_{2} \cdots i_{n}} \right| \\
	\end{align*}
	
	(2) Considering the combination computation in eq. \eqref{eq:a10} and the recursive time length $N$, the computational complexity of eq. \eqref{eq:a10} in the $N$-th recursive computation stage is less than or equal to
	\begin{equation} \label{eq:a11}
	\left(r \times r^{r/2} \right) ^2 \times \frac{ \left(rN-2 \right)!}{ \left(rN-n \right)!(n-2)!}+1
	\end{equation}
	
	\noindent times computing the $n \times n$ determinant. Then the complexity for the full recursive computation can be noted as the polynomial time $ \mathcal{O}( N^{n-1}r^{n+r})$, \textit{i.e.}, $ \mathcal{O}(N^{n-1})$ on time variable $N$. 
	\qed
	
	When $r=1$, \textit{ i.e.}, $B$ in matrix pair $ \{A,B \}$ is an $n \times 1$ vector, eq. \eqref{eq:a10} can be simplified to
	\begin{align}
		V_{n} (E_{n}(P_{N})) & = \left(1+ \left| \det A \right| \right)V_{n}(E_{n}(P_{N-1}))- \left| \det A \right|V_{n}(E_{n}(P_{N-2})) \notag \\
		& \quad+ \sum_{(i_{2},i_{3}, \cdots,i_{n-1}) \in \varTheta_{1,N-2}^{n-2}} \left| \det [B,A^{i_{2}}B, \cdots,A^{i_{n-1}}B,A^{N-1}B] \right| \label{eq:a12}
	\end{align}
	By eq. \eqref{eq:a10} or \eqref{eq:a12}, the volume of zonotope $E_n(P_N)$ can be computed recursively with the time variable $N$, and the corresponding complexity will be reduced from $ \mathcal{O}(N^{n})$ to $ \mathcal{O}(N^{n-1})$ .
	
	\section{Volume Computation for Matrix $A$ with $n$ real eigenvalues}
	
	\subsection{a lemma on the determinant of quasi-Vandermonde matrices}
	
	By \textbf{Corollary \ref{c12}}, the volume computation of the zonotope $E_{n}(P_{N})$ spanned by the general matrix pair $ \{A,B \}$ can be made, and the more effective computation methods of the zonotope volume for a matrix $A$ with $n$ real eigenvalues will be studied here.
	
	First, for that matrix pair, the following lemma about the sign of a class of quasi-Vandermonde matrices is proposed and proven.
	
	\begin{lemma} \label{l12} For any $n>0$, if $0<k_{1}<k_{2}< \cdots<k_{n}$ and $0< \lambda_{1}< \lambda_{2}< \cdots< \lambda_{n}$, we have
		\begin{equation} \label{eq:a13}
		F_{ \lambda_{1} \lambda_{2} \cdots \lambda_{n}}^{k_1,k_{2} \cdots k_{n}} = \det \left[ \begin{array}{cccc}
		\lambda_{1}^{k_{1}} & \lambda_{1}^{k_{2}} & \cdots & \lambda_{1}^{k_{n}} \\
		\lambda_{2}^{k_{1}} & \lambda_{2}^{k_{2}} & \cdots & \lambda_{2}^{k_{n}} \\
		\vdots & \vdots & \ddots & \vdots \\
		\lambda_{n}^{k_{1}} & \lambda_{n}^{k_{2}} & \cdots & \lambda_{n}^{k_{n}}
		\end{array} \right] >0
		\end{equation}
	\end{lemma}
	
	\textbf{Proof}. Let $ \alpha_{i}=
	\left[ \begin{array}{cccc}
	\lambda_{1}^{k_{i}} & \lambda_{2}^{k_{i}} & \cdots & \lambda_{n}^{k_{i}}
	\end{array} \right] ^T $, $i= \overline{1,n}$.
	In fact, $F_{ \lambda_{1} \lambda_{2} \cdots \lambda_{n}}^{k_1 k_{2} \cdots k_{n}}$ defined above is the oriented volume of the polytope spanned by the vectors
	$ \alpha_{i}(i= \overline{1,n})$
	in $n$-D space. From the representation of vectors $ \alpha_{1}$ and $ \alpha_{2}$, we know that vector $ \alpha_{2}$ can be regarded as the linear transformation result from vector $ \alpha_{1}$ via the following transformation matrix:
	
	$$
	\Lambda_1= \left [ \textrm{diag} \left \{ \lambda_{1}, \lambda_{2}, \cdots, \lambda_{n} \right \} \right ]^{k_2-k_1}
	$$
	\noindent where $ \textrm{diag} \{ \bullet \}$ denotes the diagonal matrix. Similarly, vector $ \alpha_{i}(i=3,4, \dots,n)$ can be obtained from vector $ \alpha_{i-1}$.
	
	When $ \lambda_{i}(i= \overline{1,n})$ satisfy $0< \lambda_{1}< \lambda_{2}< \cdots< \lambda_{n}$, it can be proved that after the linear transformation via matrix $ \Lambda_1$, vector $ \alpha_{1}$ to vector $ \alpha_{2}$ constitute a right-handed system and satisfy the right-hand rule. Analogously, we know that vector $ \alpha_{i}(i=3,4, \dots,n)$ is in the right-handed systems spanned by vectors $ \left \{ \alpha_{1}, \alpha_{2}, \cdots, \alpha_{i-1} \right \}$. Therefore, considering that vectors $ \left \{ \alpha_{1}, \alpha_{2}, \cdots, \alpha_{n} \right \}$ satisfy the right-handed system and are in the first quadrant of the $n$-D space,
	according to the basic theory of linear algebra, the oriented volume of the polytope spanned by these vectors must satisfy
	$$ \det \left[ \alpha_{1}, \alpha_{2}, \cdots, \alpha_{n} \right] >0 $$
	\textit{i.e.}, eq. \eqref{eq:a13} holds.
	\qed
	
	\subsection{Recursive Computation Method with Linear Time Complexity}
	
	\subsubsection{Algorithm}
	
	When the $n$ eigenvalues of the matrix $A$ are different and real, there must exist a transformation matrix $W$ such that
	\begin{align}
		\Lambda & =WAW^{-1}= \textrm{diag} \left \{ \lambda_{1}, \lambda_{2}, \cdots, \lambda_{n} \right \} \label{eq:a14} \\
		\Gamma & =WB \label{eq:a15} \\
		\overline P_N & = \left[ \Gamma, \Lambda \Gamma, \cdots \Lambda ^{N-1} \Gamma \right] = WP_N \label{eq:a16}
	\end{align}
	\noindent and then it is proven easily that for the reversible transformation matrix $W$, the volume of the zonotope $E_n( \overline P_N)$ generated by the matrix pair $( \Lambda, \Gamma)$ satisfies
	\begin{equation}
	V_n(E_n(P_N))= \left| \det W \right|^{-1} V_n(E_n( \overline P_N))
	\end{equation}
	\noindent Therefore, $V_n(E_n(P_N))$ with $n$ differentially real eigenvalues can be gotten by computing $V_n(E_n( \overline P_N))$. Later we will discuss in detail how to effectively compute $V_n(E_n( \overline P_N))$.
	
	As we know, the most practical discrete-time systems are the sampling systems from the continuous-time systems, and the eigenvalues of continuous-time systems and the corresponding sampling systems satisfy \cite{Kailath1980}
	\begin{equation}
	\lambda_{i}= \exp( \mu_{i}T) \quad i= \overline{1,n}
	\end{equation}
	\noindent where $T$ is the sampling period, and $ \mu_{i}$ and $ \lambda_{i}(i= \overline{1,n})$ are the eigenvalues of the continuous-time systems and sampling systems, respectively. Therefore, the relationships between $ \lambda_{i}$ and $ \mu_{i}$ are
	\begin{align*}
		\lambda_{i}>0 \longleftrightarrow \quad \textrm{Im}( \mu_{i})=0 \\
		\textrm{Im}( \lambda_{i}) \neq0 \longleftrightarrow \quad \textrm{Im}( \mu_{i}) \neq0
	\end{align*}
	and there exists no $ \lambda_{i} \le 0$ for finite eigenvalues $ \mu_{i}$, where $ \textrm{Im}(z)$ is the imaginary part of the complex number $z$. Hence, if the eigenvalues of the sampling systems are real, they must be positive.
	
	When matrices $B$ and $ \Gamma$ are only vectors, \textit{i.e.} the linear discrete systems \eqref{eq:a01} are with a single input, then the volumes of the zonotopes $E_n( P_N)$ and $E_n( \overline P_N)$ can be computed recursively with complexity $ \mathcal{O}(N)$, and the corresponding result can be determined by the following theorem.
	
	\begin{theorem} \label {t30} If
		$ \Lambda$ is a diagonal matrix and $ \Gamma$ is only a vector, the volume of the zonotope $E_n( \overline P_N)$ generated by matrix pair $ \{ \Lambda, \Gamma \}$ can be computed with computational complexity $ \mathcal{O}(N)$ by the following equation:
		\begin{equation}
		V_n(E_n( \overline P_N)) = \left| \prod_{i=1}^{n} \beta_{i} \right| V_{N}^{ \lambda_{1} \lambda_{2} \cdots \lambda_{n}} \label{eq:a17}
		\end{equation}
		where
		\begin{align}
			\Lambda & = \textrm{diag} \left \{ \lambda_{1}, \lambda_{2}, \cdots, \lambda_{n} \right \} \quad \quad 0< \lambda_{1}< \lambda_{2}< \cdots< \lambda_{n} \notag \\
			\Gamma & = \left[ \beta_{1}, \beta_{2}, \cdots, \beta_{n} \right]^{T} \notag \\
			V_{N}^{ \lambda_{1} \lambda_{2} \cdots \lambda_{n}}
			& = \sum_{(i_{1},i_{2}, \cdots,i_{n}) \in \Omega_{N-1}^{n}}F_{ \lambda_{1} \lambda_{2} \cdots \lambda_{n}}^{i_{1}i_{2} \cdots i_{n}} \notag \\
			& =V_{N-1}^{ \lambda_{1} \lambda_{2} \cdots \lambda_{n}}+ \sum_{j=1}^{n}(-1)^{n+j} \lambda_{j}^{N-1}V_{N-1}^{ \lambda_{1} \lambda_{2} \cdots \lambda_{n} \setminus \lambda_{j}} \label{eq:a18}
		\end{align}
		where '$ \lambda_{1} \lambda_{2} \cdots \lambda_{n} \setminus \lambda_{j}$' means that $ \lambda_{j}$ is deleted from sequence $ \lambda_{1} \lambda_{2} \cdots \lambda_{n}$.
	\end{theorem}
	
	\textbf{Proof.} (1) By \textbf{Theorem \ref{t10}} and eq. \eqref{eq:a16}, we have
	\begin{align}
		V_n(E_n( \overline P_N))
		& = \sum_{(k_{1},k_{2}, \cdots,k_{n}) \in \Omega_{N-1}^{n}} \left| \det \left[ \varLambda^{k_{1}} \Gamma, \varLambda^{k_{2}} \Gamma, \cdots, \varLambda^{k_{n}} \Gamma \right] \right| \notag \\
		& = \left| \prod_{i=1}^{n} \beta_{i} \right| \sum_{(k_{1},k_{2}, \cdots,k_{n}) \in \Omega_{N-1}^{n}} \left|F_{ \lambda_{1} \lambda_{2} \cdots \lambda_{n}}^{k_{1}k_{2} \cdots k_{n}} \right| \label{eq:a19}
	\end{align}
	
	For any $ \lambda_{i}$ and $k_{i} $ that satisfy
	$$ 0< \lambda_{1}< \lambda_{2}< \cdots< \lambda_{n} \quad \textrm{and} \quad 0 \leq k_{1}<k_{2}< \cdots<k_{n}
	$$
	\noindent according to \textbf{Lemma \ref{l12}}, eq. \eqref{eq:a19} can be rewritten as
	\begin{equation}
	V_n(E_n( \overline P_N)) = \left| \prod_{i=1}^{n} \beta_{i} \right|V_{N}^{ \lambda_{1} \lambda_{2} \cdots \lambda_{n}} \label{eq:a20}
	\end{equation}
	\noindent where
	\begin{equation} \label{eq:a21}
	V_{N}^{ \lambda_{1} \lambda_{2} \cdots \lambda_{n}}= \sum_{(i_{1},i_{2}, \cdots,i_{n}) \in \Omega_{N-1}^{n}}F_{ \lambda_{1} \lambda_{2} \cdots \lambda_{n}}^{i_{1}i_{2} \cdots i_{n}}
	\end{equation}
	\noindent Then the above $V_{N}^{ \lambda_{1} \lambda_{2} \cdots \lambda_{n}}$ can be computed recursively as follows:
	\begin{align*}
		V_{N}^{ \lambda_{1} \lambda_{2} \cdots \lambda_{n}} & = V_{N-1}^{ \lambda_{1} \lambda_{2} \cdots \lambda_{n}}+
		\sum_{ \left(i_{1},i_{2}, \cdots,i_{n-1} \right) \in \Omega_{N-2}^{n-1}}F_{ \lambda_{1} \lambda_{2} \cdots \lambda_{n-1} \lambda_{n}}^{i_{1}i_{2} \cdots i_{n-1},N-1} \notag \\
		& =V_{N-1}^{ \lambda_{1} \lambda_{2} \cdots \lambda_{n}}+ \sum_{ \left(i_{1},i_{2}, \cdots,i_{n-1} \right) \in \Omega_{N-2}^{n-1}} \det \left[ \begin{array}{ccccc}
			\lambda_{1}^{i_{1}} & \lambda_{1}^{i_{2}} & \cdots & \lambda_{1}^{i_{n-1}} & \lambda_{1}^{N-1} \\
			\lambda_{2}^{i_{1}} & \lambda_{2}^{i_{2}} & \cdots & \lambda_{2}^{i_{n-1}} & \lambda_{2}^{N-1} \\
			\vdots & \vdots & \ddots & \vdots & \vdots \\
			\lambda_{n}^{i_{1}} & \lambda_{n}^{i_{2}} & \cdots & \lambda_{n}^{i_{n-1}} & \lambda_{3}^{N-1}
		\end{array} \right]
		\notag \\
		&=V_{N-1}^{ \lambda_{1} \lambda_{2} \cdots \lambda_{n}}+ \sum_{ \left(i_{1},i_{2}, \cdots,i_{n-1} \right) \in \Omega_{N-2}^{n-1}} \sum_{j=1}^{n}(-1)^{n+j} \lambda_{j}^{N-1}F_{ \lambda_{1} \lambda_{2} \cdots \lambda_{n} \setminus \lambda_{j}}^{i_{1}i_{2} \cdots i_{n-1}}
		\notag \\
		&=V_{N-1}^{ \lambda_{1} \lambda_{2} \cdots \lambda_{n}}+ \sum_{j=1}^{n}(-1)^{n+j} \lambda_{j}^{N-1}V_{N-1}^{ \lambda_{1} \lambda_{2} \cdots \lambda_{n} \setminus \lambda_{j}}
	\end{align*}
	Therefore, eqs. \eqref{eq:a17} and \eqref{eq:a18} hold.
	
	(2) The computational complexity of the recursive eq. \eqref{eq:a18} can be divided into two parts. One is $n(N-2)$ for computing the power $ \lambda_{j}^{i}(i= \overline{1,N-1};j= \overline{1,n})$, and the other is the rest of the complexity in the recursive process. The recursive process can be described by the following array:
	$$ \begin{array}{cccccccc}
	C_{n}^{n} & n & V_{N}^{ \lambda_{1} \lambda_{2} \cdots \lambda_{n}} & \rightarrow & V_{N-1}^{ \lambda_{1} \lambda_{2} \cdots \lambda_{n}} & \rightarrow & \cdots & V_{n}^{ \lambda_{1} \lambda_{2} \cdots \lambda_{n}} \\
	& & \downarrow & & \downarrow & & \cdots \\
	C_{n}^{n-1} & n-1 & V_{N-1}^{ \lambda_{1} \lambda_{2} \cdots \lambda_{n} \setminus \lambda_{j}} & \rightarrow & V_{N-2}^{ \lambda_{1} \lambda_{2} \cdots \lambda_{n} \setminus \lambda_{j}} & \rightarrow & \cdots & V_{n-1}^{ \lambda_{1} \lambda_{2} \cdots \lambda_{n} \setminus \lambda_{j}} \\
	& & \downarrow & & \downarrow & & \cdots \\
	& & \vdots & \vdots & \vdots & \vdots & \ddots & \vdots \\
	C_{n}^{1} & 1 & V_{N-n+1}^{ \lambda_{i}} & \rightarrow & V_{N-n}^{ \lambda_{i}} & \rightarrow & \cdots & V_{1}^{ \lambda_{i}}
	\end{array}
	$$
	
	In the above array:
	
	1) Each element $C_{n}^{k}$ of the first column is the variable number of the recursive variables $V_{*}^{ \lambda_{j_{1}} \lambda_{j_{2}}, \cdots \lambda_{j_{k}}}$.
	
	2) Each element of the second column is the increasing number of multiplications to compute the recursive variable $V_{*}^{ \lambda_{j_{1}} \lambda_{j_{2}}, \cdots \lambda_{j_{k}}}$ by eq. \eqref{eq:a18}.
	
	3) Each element $V_{i}^{ \lambda_{j_{1}} \lambda_{j_{2}} \cdots \lambda_{j_{i}}}$ of the last column is essentially the value of each order Vandermonde determinant, and its complexity is $i(i-1)/2$.
	
	4) Each element $ V_{j}^{ \lambda_{i}}$ of the last row equals $ V_{j-1}^{ \lambda_{i}}+ \lambda_{i}^j$, and the complexity of the power $ \lambda_{j}^{i}$ has been computed above.
	\noindent In summary, the computational complexity $Q_{N}^{(n)}$ for $V_{N}^{ \lambda_{1} \lambda_{2} \cdots \lambda_{n}}$ is
	\begin{align}
		Q_{N}^{(n)} & =n(N-2)+ \sum_{i=2}^{n}C_{n}^{i} \times i \times(N-n)+ \sum_{i=2}^{n} \frac{i(i-1)}{2} \notag \\
		& =n(N-2)+ \sum_{i=2}^{n} \frac{n!}{(n-i)!i!} \times i \times(N-n)+ \sum_{i=2}^{n} \frac{i(i-1)}{2} \notag \\
		& = \mathcal{O}(n^{n/2+2}N) \label{eq:a22}
	\end{align}
	\textit{i.e.}, the complexity for the volume of the zonotope $E_n(P_N)$ is linear complexity $ \mathcal{O}(N)$ on the time variable $N$.
	\qed
	
	\section{Analytic Computation Method for Infinite-time $E_n(P_ \infty )$}
	
	For many analysis problems on the controllability of practical dynamic systems \eqref{eq:a01}, our focus is on the infinite-time controllable region $R_{c, \infty}$ and reachable region $R_{r, \infty}$. The computational cost of these region volumes by \textbf{Theorem \ref{t10}}, \textbf{Corollary \ref{c12}}, and \textbf{Theorem \ref{t30}} will approach infinity. We now propose a theorem on an analytic computation method with complexity $ \mathcal{O}(1)$ that has nothing to do with the time variable $N(N \rightarrow \infty)$, and we prove it as follows.
	
	\begin{theorem} \label{t31}
		For the $n$ eigenvalues $ \lambda_i(i= \overline{1,n})$ of the matrix $A$ satisfying
		$$
		0< \lambda_1< \lambda_2< \cdots< \lambda_n<1,
		$$
		the volume of the infinite-time $E_n(P_ \infty )$ is as
		\begin{equation} \label{eq:a23}
		V_{ \infty}^{ \lambda_{1} \lambda_{2} \cdots \lambda_{n}} = \Phi_{ \lambda_{1} \lambda_{2} \cdots \lambda_{n}}
		\end{equation}
		where
		\begin{equation} \label{eq:a24}
		\Phi_{ \lambda_{1} \lambda_{2} \cdots \lambda_{n}}= \left( \prod_{1 \leq j_{1}<j_{2} \leq n} \frac{ \lambda_{j_{2}}- \lambda_{j_{1}}}{1- \lambda_{j_{1}} \lambda_{j_{2}}} \right) \left( \prod_{i=1}^{n} \frac{1}{1- \lambda_{i}} \right)
		\end{equation}
	\end{theorem}
	
	\textbf{Proof}. The inductive method can be used to prove the theorem.
	
	(1) When $n=1$, we have
	$$
	V_{ \infty}^{ \lambda_{1}} = \sum_{i_{1}=0}^{ \infty}F_{ \lambda_{1}}^{i_{1}}= \frac{1}{1- \lambda_{1}}
	$$
	\noindent \textit and eq. \eqref{eq:a23} holds.
	
	(2) Assume that for all $n=s \leq m-1$, eq. \eqref{eq:a23} holds, \textit{i.e.},
	\begin{equation} \label{eq:a25}
	V_{ \infty}^{ \lambda_{1} \lambda_{2} \cdots \lambda_{s}} = \Phi_{ \lambda_{1} \lambda_{2} \cdots \lambda_{s}}
	\end{equation}
	
	In addition, according to definition \eqref{eq:a21} of $V_{N}^{ \lambda_{1} \lambda_{2} \cdots \lambda_{s}}$, we have
	\begin{align}
		V_{ \infty}^{ \lambda_{1} \lambda_{2} \cdots \lambda_{s}}
		& = \sum_{(k_{1},k_{2}, \cdots,k_{s}) \in \Omega_{0, \infty}^{s}}F_{ \lambda_{1} \lambda_{2} \cdots \lambda_{s}}^{k_{1}k_{2} \cdots k_{s}} \notag \\
		& = \sum_{(k_{2}, \cdots,k_{s}) \in \Omega_{1, \infty}^{s-1}}F_{ \lambda_{1} \lambda_{2} \cdots \lambda_{s}}^{0,k_{2} \cdots k_{s}}+ \sum_{(k_{1},k_{2,} \cdots,k_{s}) \in \Omega_{1, \infty}^{s}}F_{ \lambda_{1} \lambda_{2} \cdots \lambda_{s}}^{k_{1}k_{2} \cdots k_{s}} \notag \\
		&
		= \sum_{(k_{2}, \cdots,k_{s}) \in \Omega_{1, \infty}^{s-1}} \sum_{k=1}^{s}(-1)^{1+k}F_{ \lambda_{1} \lambda_{2} \cdots \lambda_{s} \setminus \lambda_{k}}^{k_{2} \cdots k_{s}} \notag \\
		& \quad + \Upsilon_s \sum_{(k_{1},k_{2,} \cdots,k_{s}) \in \Omega_{0, \infty}^{s}}F_{ \lambda_{1} \lambda_{2} \cdots \lambda_{s}}^{k_{1}k_{2} \cdots k_{s}} \notag \\
		&
		= \sum_{(k_{2}, \cdots,k_{s}) \in \Omega_{0, \infty}^{s-1}} \sum_{k=1}^{s}(-1)^{1+k} \Upsilon_{s \setminus k} F_{ \lambda_{1} \lambda_{2} \cdots \lambda_{s} \setminus \lambda_{k}}^{k_{2} \cdots k_{s}} \notag \\
		& \quad + \Upsilon_s V_{ \infty}^{ \lambda_{1}, \lambda_{2}, \cdots, \lambda_{s}} \notag \\
		&
		= \sum_{k=1}^{s}(-1)^{1+k} \Upsilon_{s \setminus k} V_{ \infty}^{ \lambda_{1} \lambda_{2} \cdots \lambda_{s} \setminus \lambda_{k}}+ \Upsilon_s V_{ \infty}^{ \lambda_{1} \lambda_{2} \cdots \lambda_{s}} \label{eq:a26}
	\end{align}
	\noindent \textit{i.e.},
	\begin{equation} \label{eq:a27}
	\left(1- \Upsilon_s \right)V_{ \infty}^{ \lambda_{1} \lambda_{2} \cdots \lambda_{s}} = \sum_{k=1}^{s}(-1)^{1+k} \Upsilon_{s \setminus k} V_{ \infty}^{ \lambda_{1} \lambda_{2} \cdots \lambda_{s} \setminus \lambda_{k}}
	\end{equation}
	where
	$$
	\Upsilon_s = \prod_{i=1}^{s} \lambda_{i} , \quad \Upsilon_{s \setminus k} = \prod_{i={ \overline {1,s} \setminus k}} \lambda_{i}
	$$
	
	(3) Next, it will be proved that for $n=m$, eq. \eqref{eq:a23} holds.
	
	Similar to the proof of eq. \eqref{eq:a27}, considering the above assumption for $V_{ \infty}^{ \lambda_{1} \lambda_{2} \cdots \lambda_{s}}$ with $s \le m-1$, we have
	\begin{align}
		\left(1- \Upsilon_n \right) V_{ \infty}^{ \lambda_{1} \lambda_{2} \cdots \lambda_{n}}
		& = \sum_{k=1}^{n}(-1)^{1+k} \Upsilon_{n \setminus k} V_{ \infty}^{ \lambda_{1} \lambda_{2} \cdots \lambda_{n} \setminus \lambda_{k}} \notag \\
		& = \sum_{k=1}^{n}(-1)^{1+k} \Upsilon_{n \setminus k} \Phi_{ \lambda_{1} \lambda_{2} \cdots \lambda_{n} \setminus \lambda_{k}} \label{eq:a28}
	\end{align}
	\noindent By the definition of $V_{N}^{ \lambda_{1} \lambda_{2} \cdots \lambda_{n}}$, it can be proved that for any $ \lambda_{j_{1}}$ and $ \lambda_{j_{2}}, \lambda_{j_{2}}- \lambda_{j_{1}}$ is a factor of $V_{ \infty}^{ \lambda_{1} \lambda_{2} \cdots \lambda_{n}}$. And then,
	by the definition of $ \Phi_{ \lambda_{1} \lambda_{2} \cdots \lambda_{n} }$,
	$V_{ \infty}^{ \lambda_{1} \lambda_{2} \cdots \lambda_{n}}$ can be represented as
	\begin{equation} \label{eq:a29}
	V_{ \infty}^{ \lambda_{1} \lambda_{2} \cdots \lambda_{n}} =H_{ \lambda_{1} \lambda_{2} \cdots \lambda_{n}} \Phi_{ \lambda_{1} \lambda_{2} \cdots \lambda_{n} }
	\end{equation}
	\noindent where $H_{ \lambda_{1} \lambda_{2} \cdots \lambda_{n}}$ is an undetermined polynomial function on $ \lambda_{i}(i= \overline{1,n})$. By eqs. \eqref{eq:a28} and \eqref{eq:a29}, it can be proved that the highest order of $H_{ \lambda_{1} \lambda_{2} \cdots \lambda_{n}}$ is $n-1$, \textit{i.e.}, the polynomial function $H_{ \lambda_{1} \lambda_{2} \cdots \lambda_{n}}$ can be described as
	\begin{equation} \label{eq:a30}
	H_{ \lambda_{1} \lambda_{2} \cdots \lambda_{n}}= \sum_{i=0}^{n-1} \sum_{ \sum_{k=1}^{n}j_{k}=i}c_{j_{1}j_{2} \cdots j_{n}}^{(n,i)} \lambda_{1}^{j_{1}} \lambda_{2}^{j_{2}} \cdots \lambda_{n}^{j_{n}}
	\end{equation}
	\noindent where $c_{j_{1}j_{2} \cdots j_{n}}^{(n,i)}$ is an undetermined coefficient. Because of the symmetry of $H_{ \lambda_{1} \lambda_{2} \cdots \lambda_{n}}$ on $ \lambda_{i}(i= \overline{1,n})$, we have
	\begin{equation} \label{eq:a31}
	c_{j_{1}j_{2} \cdots j_{n}}^{(n,i)}=c_{k_{1}k_{2} \cdots k_{n}}^{(n,i)}
	\end{equation}
	\noindent where $ \{k_{1},k_{2}, \cdots,k_{n} \}$ is any other permutation of $ \{j_{1},j_{2}, \cdots,j_{n} \}$.
	
	Therefore, by eqs. \eqref{eq:a28} and \eqref{eq:a29}, we know that to prove eq. \eqref{eq:a23} is true is equivalent to proving the following equation is true:
	\begin{align}
		H_{ \lambda_{1} \lambda_{2} \cdots \lambda_{n}} \left(1- \Upsilon_n \right)
		\Phi_{ \lambda_{1} \lambda_{2} \cdots \lambda_{n} }
		= \sum_{k=1}^{n}(-1)^{1+k} \Upsilon_{n \setminus k} \Phi_{ \lambda_{1} \lambda_{2} \cdots \lambda_{n} \setminus \lambda_{k} } \label{eq:a32}
	\end{align}
	Next, eq. \eqref{eq:a32} will be proved true for any real variables ,$ \lambda_i(i= \overline{1,n})$ and then eq. \eqref{eq:a23} must be true for all $ \lambda_i \in (0,1)$.
	
	After playing by $1- \lambda_{n}$ and then letting $ \lambda_{n}=1$, considering that $ \Phi_{ \lambda_{1} \lambda_{2} \cdots \lambda_{n-1},1 }= \Phi_{ \lambda_{1} \lambda_{2} \cdots \lambda_{n-1} }$, the two sides in the above equation can be rewritten as follows.
	\begin{align}
		\text{left side} & = H_{ \lambda_{1} \lambda_{2} \cdots \lambda_{n-1},1} \left(1- \Upsilon_{n-1} \right) \Phi_{ \lambda_{1} \lambda_{2} \cdots \lambda_{n-1} } \notag \\
		& =H_{ \lambda_{1} \lambda_{2} \cdots \lambda_{n-1},1} \left(1- \Upsilon_{n-1} \right)V_{ \infty}^{ \lambda_{1} \lambda_{2} \cdots \lambda_{n-1}} \label{eq:a33} \\
		\text{right side} & = \sum_{k=1}^{n-1}(-1)^{1+k}
		\Upsilon_{n-1 \setminus k}
		\Phi_{ \lambda_{1} \lambda_{2} \cdots \lambda_{n-1} \setminus \lambda_{k} } \notag \\
		& = \sum_{k=1}^{n-1}(-1)^{1+k} \Upsilon_{n-1 \setminus k} V_{ \infty}^{ \lambda_{1} \lambda_{2} \cdots \lambda_{n-1} \setminus \lambda_{k}} \label{eq:a34}
	\end{align}
	
	Therefore, by eq. \eqref{eq:a32}, we have
	\begin{equation} \label{eq:a35}
	H_{ \lambda_{1} \lambda_{2} \cdots \lambda_{n-1},1}=1
	\end{equation}
	\noindent \textit{i.e.},
	\begin{equation} \label{eq:a36}
	H_{ \lambda_{1} \lambda_{2} \cdots \lambda_{n-1},1}= \sum_{i=0}^{n-1} \sum_{ \sum_{k=1}^{n}j_{k}=i}c_{j_{1}j_{2} \cdots j_{n-1},0}^{(n,i)} \lambda_{1}^{j_{1}} \lambda_{2}^{j_{2}} \cdots \lambda_{n-1}^{j_{n-1}}=1
	\end{equation}
	\noindent Then, when $ \lambda_{i}=0(i= \overline{1,n-1})$, we can get
	\begin{equation} \label{eq:a37}
	c_{0,0, \cdots,0}^{(n,0)}=1
	\end{equation}
	
	In addition, for some $i \in[1,n-1]$ and $ \left \{ j_{1},j_{2}, \cdots,j_{n-1} \right \}$ in eq. \eqref{eq:a36}, if
	\begin{equation}
	c_{j_{1}j_{2} \cdots j_{n-1},0}^{(n,i)} \neq0,
	\end{equation}
	\noindent then the dimension of the real solution space of variables $ \lambda_{i}(i= \overline{1,n-1})$ in eq. \eqref{eq:a36} is less than or equal to $n-2$. But, because the equation holds for any $ \lambda_{i} \left(i= \overline{1,n-1} \right)$, the dimension of the real solution space must be $n-1$. The above two conclusions contradict each other, and so it can be proved that
	\begin{equation}
	c_{j_{1}j_{2} \cdots j_{n-1},0}^{(n,i)}=0 \quad \sum_{k=1}^{n}j_{k}=i, \forall i \in[1,n-1]
	\end{equation}
	\noindent Then, by \eqref{eq:a31}, for all $i \in[1,n-1]$ and $j_{k}(k= \overline{1,n})$ with $ \sum_{k=1}^{n}j_{k}=i$, we have
	\begin{equation}
	c_{j_{1}j_{2} \cdots j_{n}}^{(n,i)}=0 \quad \exists j_{i}=0
	\end{equation}
	
	Because $ \sum_{k=1}^{n}j_{k}=i$ and $i<n$, it is sure that at least one element of $ \{j_{1},j_{2}, \cdots,j_{n} \}$ is zero. Therefore, for any $c_{j_{1}j_{2} \cdots j_{n}}^{(n,i)}$, we have
	\begin{equation}
	c_{j_{1}j_{2} \cdots j_{n}}^{(n,i)}=0 \quad \sum_{k=1}^{n}j_{k}=i,i \in[1,n-1]
	\end{equation}
	\noindent \textit{i.e.},
	\begin{equation}
	H_{ \lambda_{1} \lambda_{2} \cdots \lambda_{n}}=1
	\end{equation}
	
	Thus, by eq. \eqref{eq:a29}, we know that when $m=n$, eq. \eqref{eq:a23} is also true.
	
	In summary, eq. \eqref{eq:a23} is proved to be true by the inductive method.
	\qed
	
	For the volume computation of the infinite-time zonotope $E_n(P_ \infty )$, the computational complexity of eq. \eqref{eq:a23} is $ \mathcal{O}(n^2)$, and it has nothing to do with the time variable $N(N \rightarrow \infty )$.

	\section{Numerical Experiments }
	
	In this section, two numerical experiments for volume computation of the controllable and reachable regions are carried out.
	
	\begin{example} \label{e01} Computing the volume of the finite-time reachable region of the following linear discrete-time system:
		\begin{equation} \label{eq:a38}
		x_{k+1}= \left[ \begin{array}{ccc}
		0 & 1 & 0 \\
		0 & 0 & 1 \\
		0.9596 & -2.9196 & 2.96
		\end{array} \right]x_{k}+ \left[ \begin{array}{c}
		0 \\
		0 \\
		
		1
		\end{array} \right]u_{k}
		\end{equation}	
	\end{example}
	
	The three eigenvalues of the matrix $A$ are \{ 0.9517 1.0000 1.0083 \}, and the volume of the finite-time reachable region can be computed by
	\textbf{Theorem \ref{t10}}, \textbf{Corollary \ref{c12}}, and \textbf{Theorem \ref{t30}}. The numerical experiments are carried out with the Intel Core i7-7700 3.6GHz CPU and MATLAB R2012a. The computational results are shown in Table \ref{tab:1}, where $N$ is the number of sampling steps, $v_r$ is the region volume, $n_d$ is the number of times computing the $n \times n$ determinants, $C_t$ is the computational time cost, and $n_p$ is the number of multiplications only in recursive equations \eqref{eq:a17} and \eqref{eq:a18}. From the table, we can see that the volumes computed by the three methods are exactly the same, but the computation methods proposed in \textbf{Corollary \ref{c12}} and \textbf{Theorem \ref{t30}} can greatly reduce the computational complexity.
	
	\begin{table}
		\caption{Numerical results for the reachable regions}
		\label{tab:1} % Give a unique label
		\begin{tabular}{cccccccc}
			\hline \noalign{ \smallskip}
			& & \multicolumn{2}{c}{ \textbf{Theorem \ref{t10}}} & \multicolumn{2}{c}{ \textbf{Corollary \ref{c12}}} & \multicolumn{2}{c}{ \textbf{Theorem \ref{t30}}} \\
			$N$ & $v_r$ & $n_d$ & $C_t$($s$) & $n_d$ & $C_t$($s$) & $n_p$ & $C_t$($s$) \\
			\hline
			\noalign{ \smallskip}
			100 & 4.622E9 & 1.617E5 & 6.887E-1 & 4.852E3 & 2.953E-2 & 1.470E3 & 1.334E-2 \\
			200 & 1.162E11 & 1.313E6 & 5.427 & 1.970E4 & 1.086E-1 & 2.970E3 & 1.615E-2 \\
			300 & 8.015E11 & 4.455E6 & 1.857E1 & 4.455E4 & 2.395E-1 & 4.470E3 & 1.872E-2 \\
			400 & 3.553E12 & 1.059E7 & 4.373E1 & 7.940E4 & 4.219E-1 & 5.970E3 & 2.127E-2 \\
			500 & 1.274E13 & 2.071E7 & 8.519E1 & 1.243E5 & 6.586E-1 & 7.470E3 & 2.383E-2 \\
			600 & 4.057E13 & 3.582E7 & 1.487E2 & 1.791E5 & 9.508E-1 & 8.970E3 & 2.638E-2 \\
			700 & 1.199E14 & 5.692E7 & 2.370E2 & 2.440E5 & 1.299 & 1.047E4 & 2.908E-2 \\
			800 & 3.373E14 & 8.501E7 & 3.585E2 & 3.188E5 & 1.692 & 1.197E4 & 3.166E-2 \\
			\noalign{ \smallskip} \hline
		\end{tabular}
		
	\end{table}
	
	\begin{example} \label{e02} Computing the volume of the finite- and infinite-time controllable region of the following linear discrete-time system
		\begin{equation} \label{eq:a39}
		x_{k+1}= \left[ \begin{array}{cccc}
		0 & 1 & 0 & 0 \\
		0 & 0 & 1 & 0 \\
		0 & 0 & 0 & 1 \\
		-1.5629 & 5.6007 & -7.5179 & 4.48
		\end{array} \right]x_{k}+ \left[ \begin{array}{c}
		0 \\
		0 \\
		0 \\
		1
		\end{array} \right]u_{k}
		\end{equation}
	\end{example}
	
	The four eigenvalues of matrix $A$ are \{1.2049 1.1589 1.0755 1.0407 \}, the volume of the finite- and infinite-time controllable regions can be computed by \textbf{Theorem \ref{t10}}, \textbf{Corollary \ref{c12}}, \textbf{Theorem \ref{t30}}, and \textbf{Theorem \ref{t31}},
	%Eqs. \eqref{eq:a06}, \eqref{eq:a10}, \eqref{eq:a18}, and \eqref{eq:a23},
	respectively, and the experimental tools are as in \textit{Example \ref{e01}}. The numerical results shown in Table \ref{tab:2} are for the volume computation of the finite-time controllable region by \textbf{Theorem \ref{t10}}, \textbf{Corollary \ref{c12}}, and \textbf{Theorem \ref{t30}}, and the numerical results shown in Table \ref{tab:3} are for the volume computation of the infinite-time controllable region by \textbf{Theorem \ref{t31}}, where $n_{inf}$ is the number of multiplications only in eqs. \eqref{eq:a17} and \eqref{eq:a24}. These results show the effectiveness of the computational methods proposed in this paper.
	
	\begin{table}
		\caption{Numerical results for the finite-time controllable regions}
		\label{tab:2} % Give a unique label
		\begin{tabular}{cccccccc}
			\hline \noalign{ \smallskip}
			& & \multicolumn{2}{c}{ \textbf{Theorem \ref{t10}}} & \multicolumn{2}{c}{ \textbf{Corollary \ref{c12}}} & \multicolumn{2}{c}{ \textbf{Theorem \ref{t30}}} \\
			$N$ & $v_r$ & $n_d$ & $C_t$($s$) & $n_d$ & $C_t$($s$) & $n_p$ & $C_t$($s$) \\
			\hline
			\noalign{ \smallskip}
			50 & 2.388E8 & 2.303E5 & 8.134E-1 & 1.843E4 & 6.650E-2 & 1.696E3 & 1.323E-2 \\
			100 & 7.495E8 & 3.921E6 & 1.376E1 & 1.569E5 & 5.336E-1 & 3.496E3 & 1.915E-2 \\
			150 & 8.671E8 & 2.026E7 & 7.051E1 & 5.403E5 & 1.819 & 5.296E3 & 2.497E-2 \\
			200 & 8.846E8 & 6.468E7 & 2.251E2 & 1.294E6 & 4.366 & 7.096E3 & 3.080E-2 \\
			250 & 8.871E8 & 1.589E8 & 5.523E2 & 2.542E6 & 8.567 & 8.896E3 & 3.663E-2 \\
			300 & 8.874E8 & 3.308E8 & 1.154E3 & 4.411E6 & 1.485E1 & 1.070E4 & 4.245E-2 \\
			350 & 8.874E8 & 6.146E8 & 2.136E3 & 7.024E6 & 2.359E1 & 1.250E4 & 4.823E-2 \\
			400 & 8.874E8 & 1.051E9 & 3.646E3 & 1.051E7 & 3.532E1 & 1.430E4 & 5.405E-2 \\
			\noalign{ \smallskip} \hline
		\end{tabular}
	\end{table}
	
	\begin{table}
		\caption{Numerical results for the infinite-time controllable regions by Theorem \ref{t31}}
		\label{tab:3} % Give a unique label
		\begin{tabular}{cccccccc}
			\hline \noalign{ \smallskip}
			$v_r$ & $n_{inf}$  & $C_t$($s$) \\
			\hline
			\noalign{ \smallskip}
			8.874E8 & 26 & 1.328E-3 \\
			\noalign{ \smallskip} \hline
			& \end{tabular}
	\end{table}
	
	\section{Conclusions}
	In this article, we define a class of special zonotopes
	generated by matrix pair $\{A,B \}$ with finite-interval
	parameters, and then some effective computation methods with low computational complexity for the matrix $A$ with three eigenvalue-distribution cases: any  $n$  eigenvalues $\lambda_i$, $n$ differential eigenvalues  $\lambda_i \geq 0$, and $n$ different eigenvalues  $\lambda_i \in [0,1)$.
	Effective computation methods for the zonotope $E_q(P_N)$, where  the matrix $A$ has more complex eigenvalue-distribution cases, such as complex eigenvalues and repeated eigenvalues, will be investigated in our future work.
	
	\bibliographystyle{model1b-num-names}
	\bibliography{zzz}
	
\end{document}